\documentclass[cits]{PoS}

\usepackage{amsmath}
\usepackage{amssymb}
\usepackage{subfigure}
\usepackage{rotating}
\usepackage{wrapfig}

\newcommand{\newc}{\newcommand}
\newc{\bbbar}{b\bar{b}}
\newc{\Wp}{W^+}
\newc{\Wm}{W^-}
\newc{\Zgam}{Z(\gamma^*)}
\newc{\Wpm}{W^{\pm}}
 
\title{The GS09 double parton distribution functions}
 
\ShortTitle{The GS09 double parton distribution functions}
 
\author{\speaker{Jonathan Gaunt}%
         \thanks{This report summarises work performed in collaboration with C.-H. Kom, A.~Kulesza,
                  and W.~J. Stirling. JG is supported by the UK Science and Technology Facilities Council.}\\
        University of Cambridge\\
        E-mail: \email{gaunt@hep.phy.cam.ac.uk}}

\abstract{It is anticipated that hard double parton scattering (DPS) will occur
frequently in the collisions of the LHC, producing interesting signals and
significant backgrounds to certain single scattering processes. In order to
make theoretical predictions of double scattering rates and properties, the 
double parton distributions (dPDFs) $D_p^{j_1j_2}(x_1,x_2;Q_A,Q_B)$ are 
required. We discuss the first publicly available set of equal-scale ($Q_A=Q_B$)
leading order dPDFs -- the GS09 dPDFs. A brief account is given describing
how pQCD evolution effects and sum rule constraints (the latter derived by
us) have been incorporated into this set of dPDFs. We then present a summary
of a phenomenological investigation into same-sign W pair production 
conducted using GS09. In this, the DPS signal produced using GS09 is compared
with that obtained using simple products of single PDFs $\times (1-x_1-x_2)^n$,
and the single scattering backgrounds ($\Wpm\Wpm jj$, di-boson and heavy flavour)
are carefully calculated. It is found
that the correlations in GS09 manifest themselves in non-trivial kinematic
correlations between the W bosons. However, it is unlikely that these correlations
will be measurable at the LHC in the near future owing to the fact that the
background is significant even after cuts.
}
 
\FullConference{XVIII International Workshop on Deep-Inelastic Scattering and Related Subjects\\
		 April 19 -23, 2010\\
		 Convitto della Calza, Firenze, Italy\\
\vspace{10pt}
		 $\textnormal{Cavendish--HEP--10/11}$
}
 
\begin{document}
 
\section{Introduction}

\begin{wrapfigure}{r}{6cm}
\centering
\includegraphics[trim = 2cm 0.5cm 0 1cm, scale=0.7]{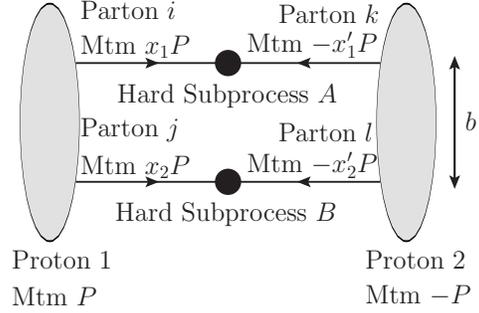}
\caption{\label{fig:schem} Schematic representation of DPS.}
\end{wrapfigure}

Double parton scattering (DPS) is said to occur when two pairs of partons participate in hard 
interactions during a single proton-proton collision. This process is drawn schematically
in figure \ref{fig:schem}, in which all of the variables relevant to DPS have also been defined.
Assuming the factorisation of the hard subprocesses $A$ and $B$, and further assuming that the 
generalised double parton distributions $\Gamma_{ij}(x_1,x_2,b;Q_A,Q_B)$ may be decomposed into a
longitudinal piece and a piece depending {\em only} on the transverse separation $b$, we may 
write the cross section for DPS as follows:
\begin{align}
\sigma^{DPS}_{(A,B)} &=& \dfrac{m}{2\sigma_{\rm eff}} \sum_{i,j,k,l}
\int dx_1dx_2dx_1'dx_2' \; D^{ij}_p(x_1,x_2;Q_A,Q_B)\;
D^{kl}_p(x_1',x_2';Q_A,Q_B)
\hat{\sigma}^A_{ik}(x_1,x_1')\hat{\sigma}^B_{jl}(x_2,x_2')
\end{align}

\vspace{-10pt}

The function $D^{ij}_p(x_1,x_2;Q_A,Q_B)$ is known as a double parton distribution function (dPDF) 
and may be interpreted at LO as the probability of finding a pair of partons in the 
proton with flavours $i$ and $j$, longitudinal momentum fractions $x_1$ and $x_2$, when probed at scales
$Q_A$ and $Q_B$. The quantity $\sigma_{eff}$ is a non-perturbative factor which is 
related to the pair density distribution in transverse space. We take it to be a constant and 
equal to the CDF measured value of 14.5mb, with the caveat that the value of this quantity should
be re-established at LHC scales using benchmark processes. $m$ is a symmetry factor that equals $1$
if $A=B$ and $2$ otherwise, and the $\hat{\sigma}^A_{ik}$ are parton-level
cross sections. 

In many of the extant studies of DPS (see \cite{Gaunt:2009re} for a complete list of references),
 a further assumption which has been applied is to take the
dPDFs as products of sPDFs (i.e. ignore longitudinal momentum correlations). It is argued that
this is justifiable at low $x_i$ due to the large population of partons at these $x$ values, and
there is some evidence from CDF \cite{Abe:1997xk} that it holds for sea partons at low $x$. Under this additional
approximation, the DPS cross section reduces to a product of single parton scattering (SPS) cross
sections:
\begin{equation}
\sigma^{DPS}_{(A,B)} =\dfrac{m}{2}
\dfrac{\sigma^S_{(A)}\sigma^S_{(B)}}{\sigma_{\rm eff}}
\label{eq:naivefact}
\end{equation}

We recall that the SPS cross section for a given hard process grows with energy as the single PDFs 
(sPDFs) are probed at lower $x$ values where they are larger. Equation \eqref{eq:naivefact} demonstrates that the DPS cross 
section goes approximately as the product of SPS cross sections -- therefore DPS cross sections grow 
faster with energy than SPS ones, and DPS will be more important at the LHC than at any previous collider. It has
already been established using \eqref{eq:naivefact} that DPS processes constitute important backgrounds
to Higgs and other interesting signals (see e.g.\ \cite{DelFabbro:1999tf}). In addition to this, DPS can 
be considered as an interesting signal process in its own right, as it reveals information about the correlations between pairs of 
partons in the proton.

For these reasons, we need to ensure we have a good theoretical handle on DPS. It has already been 
established that the form \eqref{eq:naivefact} is not adequate. In \cite{Kirschner:1979im}
a LO `double DGLAP' (dDGLAP) equation was derived describing the scaling violations of the dPDFs with a 
common factorisation scale ($Q_A=Q_B \equiv Q$). A crucial prediction of this is that even if the dPDFs may be
taken to be equal to products of sPDFs at one scale, then at any other scale the dPDFs will deviate from
factorised forms. 

Further to this, simple factorised forms do not obey the following sum rule constraints, which we have shown are
preserved by LO dDGLAP evolution if they hold at some starting scale \cite{Gaunt:2009re}:
\begin{align} \label{mtmsum}
 \text{ Momentum Sum Rule:} &&& \sum_{j_1}\int_0^{1-x_2} dx_1 x_1 D^{{j_1}{j_2}}_p(x_1,x_2;Q) = (1-x_2)D_p^{j_2}(x_2;Q)\\
 \text{ Number Sum Rule:} &&& \int_0^{1-x_2}dx_1D_p^{j_{1v}j_2}(x_1,x_2;Q)= \begin{cases}
N_{j_{1v}} D_p^{j_2}(x_2;Q) & \text{when $j_2 \ne j_1$ or $\overline{j}_1$} \\
(N_{j_{1v}}-1) D_p^{j_2}(x_2;Q) & \text{when $j_2 = j_1$} \\
(N_{j_{1v}}+1) D_p^{j_2}(x_2;Q) & \text{when $j_2 = \overline{j}_1$ \; (1.4)}
\end{cases}
\nonumber
\end{align}
The symbol $j_{1v} \equiv j_1-\overline{j}_1$ ($j_1\ne g$), and $N_{j_{1v}}$ is the
number of `valence' $j_1$ quarks in the proton. These sum rules have simple interpretations in terms of 
conditional probabilities. The first states that if you observe a parton with momentum fraction $x_2$ in
the proton, the momentum fractions of all other partons must add up to $1-x_2$. The second states 
that if you observe a parton with flavour $j$ in the proton, the number of partons of flavour $j$ elsewhere
in the proton must be reduced by one (we use the term `number effects' to describe this simple 
phenomenon).

In this contribution we discuss the first publicly available set of LO equal-scale dPDFs -- the GS09 dPDFs -- 
which incorporates pQCD evolution effects plus the sum rule constraints. Section 2 describes how the  
dPDFs were constructed incorporating these features \cite{Gaunt:2009re}, whilst section 3 summarises a 
phenomenological study of DPS performed using GS09 \cite{Gaunt:2010pi}.

\section{The GS09 dPDFs}

The GS09 dPDF package comprises a grid of dPDF values spanning the ranges $10^{-6} < x_1 < 1$, 
$10^{-6} < x_2 < 1$, $1 \text{~GeV}^2 < Q^2 < 10^9 \text{~GeV}^2$, which is available along
with interpolation code from HepForge \cite{HepForgePage}. It has been obtained by constructing
inputs that approximately satisfy the sum rules at $Q_0 = 1 \rm{~GeV}$, and then numerically
evolving these inputs up to higher scales according to the dDGLAP equation. The sPDF set to 
which we have chosen our dPDF set to correspond is (a slightly modified version of) the MSTW2008LO 
set \cite{Martin:2009iq}.

Given the paucity of experimental data regarding the dPDFs, and in accordance with simple arguments
and the CDF results, we base our inputs on factorised products of MSTW2008LO sPDFs. However, we
modify these basic forms in several ways to ensure that the input dPDFs approximately satisfy the 
sum rules. 

First, all of the dPDFs are multiplied by a factor $\rho^{ij}(x_1,x_2)$ which is designed to take
account of phase space effects. This factor should ensure the appropriate behaviour of the dPDFs near
the kinematic boundary $x_1+x_2=1$ -- namely, a smooth decrease to zero.
It was discovered that the following form for $\rho^{ij}$ gives rise
to inputs which satisfy the momentum sum rules (plus appropriate number sum rules) well:
\begin{align}
\rho^{ij}(x_1,x_2)=(1-x_1-x_2)^2(1-x_1)^{-2-\alpha(j)}(1-x_2)^{-2-\alpha(i)} 
\end{align}
$\alpha(i)$ is $0$ if $i$ is a sea parton, and $0.5$ if it is a valence parton.

We recall that there are only a finite number of valence quarks in the proton, as opposed to
an infinite number of sea quarks and gluons. Number effects are therefore
most significant in the context of valence quarks, and on this basis we have chosen to only 
take account of valence number effects in our inputs. This is done by dividing the $u_vu_v$ part of any
dPDF by two, and completely subtracting the $d_vd_v$ part. The reasoning behind
this is that removing one up valence quark essentially halves the probability 
to find another, whilst there is no chance of finding two valence down quarks
in the proton.

Finally, we have added extra terms to input distributions whose flavour indices contain $j\bar{j}$ 
combinations to take account of so-called `$j\bar{j}$ correlations'. These are essentially related 
to sea parton number effects, although they can alternatively be thought of as arising during 
evolution from some lower scale to $Q_0$ via $g \to j\bar{j}$ splittings. It is important to 
include these terms in the equal flavour valence-valence ($j_vj_v$) inputs since the
`$j\bar{j}$ correlation' term is much larger for these than the quasi-factorised piece at low $x$.

With these adjustments, our dPDF inputs satisfy all sum rules to better than $25\%$ accuracy for 
$x \lesssim 0.8$ (in the normal `double human' basis). 

\section{Effects of using GS09 dPDFs on same-sign WW DPS signal}

It is interesting to ask how the inclusion of proper pQCD evolution and sum rule constraints in 
GS09 affects experimentally measurable DPS signals. In \cite{Gaunt:2010pi}, the same-sign WW
DPS signal produced using GS09 was compared with that arising from simple factorised forms. The
factorised forms used were simple products of MSTW2008LO dPDFs multiplied by $(1-x_1-x_2)^n$, 
$n=0,1,2$ (the `MSTW$_n$' dPDFs). Same-sign WW production was chosen as the DPS process because
it has been traditionally considered as a clean channel for observation of DPS. The cross
section for same-sign WW production via SPS is suppressed to the same order of magnitude as the 
DPS cross section due to the large number of vertices required in the Feynman diagrams. What is 
more, this SPS background must always produce two jets in addition to the WW pair -- so it can
be efficiently removed via a jet veto. 

We chose to study the signal in the case when both Ws decay leptonically. A difference in the 
GS09 and MSTW$_n$ sets can be observed in the following quantity which is particularly sensitive 
to longitudinal correlations:
\begin{align}
a_{\eta_l}=\frac{\sigma(\eta_{l_1}\times\eta_{l_2}<0)-\sigma(\eta_{l_1}\times\eta_{l_2}>0)}{\sigma(\eta_{l_1}
  \times\eta_{l_2}<0)+\sigma(\eta_{l_1}\times\eta_{l_2}>0)}
\end{align}

\begin{figure}[!ht]
\setlength{\abovedisplayskip}{-50pt}
\setlength{\belowcaptionskip}{-20pt}
  \begin{center}
    \begin{tabular}{cc}
      \subfigure[Positively charged leptons]{
        \scalebox{0.6}{
          \includegraphics{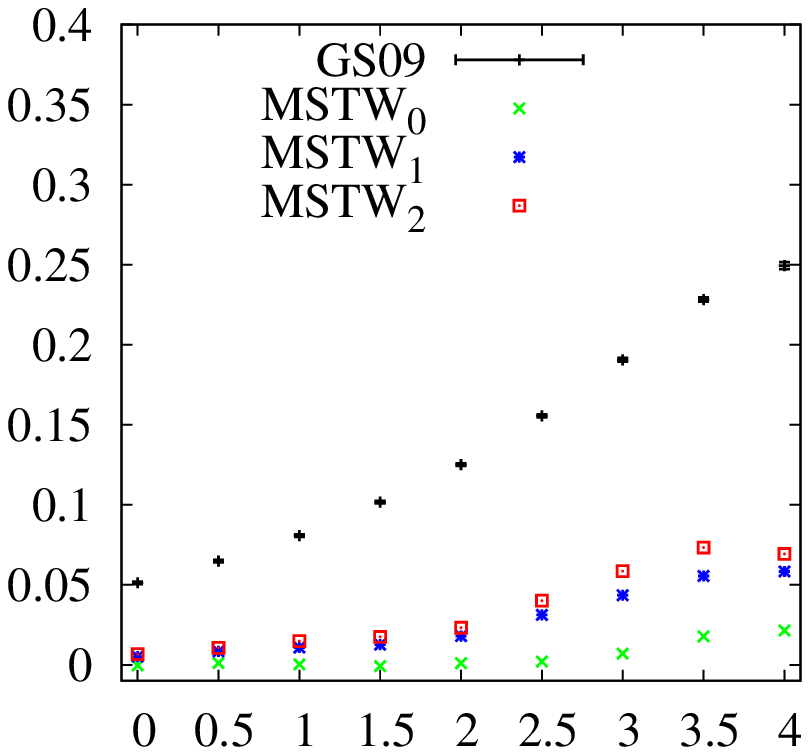}
          \label{fig:AyLtot_vs_dpdf+}
        }
        \put(-175,70){\rotatebox{90}{$a_{\eta_l}$}}
        \put(-80,-5){$\eta_l^{\textrm{min}}$}
        \put(-80,-15){$$}
      }
      &
      \subfigure[Negatively charged leptons]{
        \scalebox{0.6}{
          \includegraphics{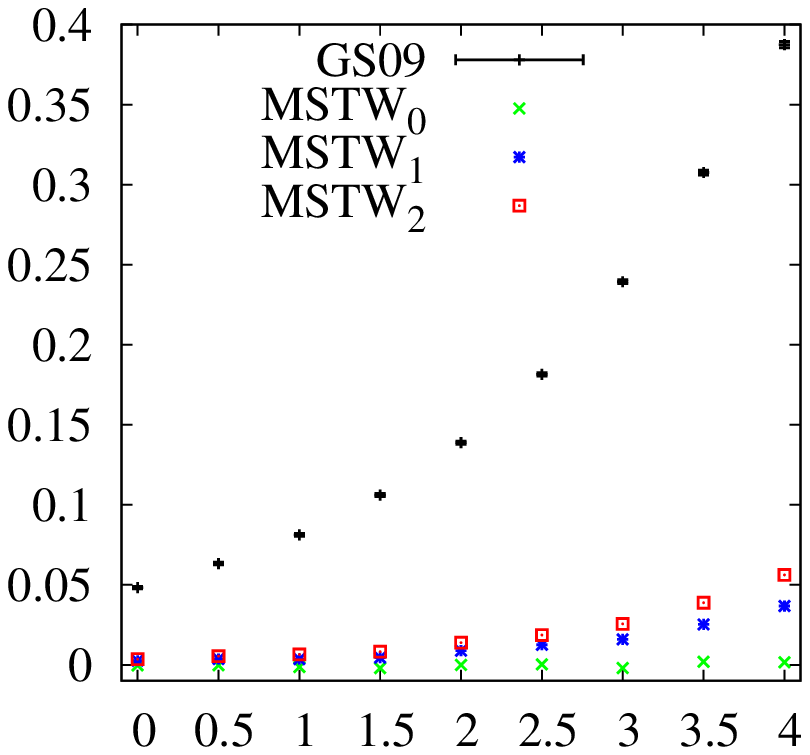}
          \label{fig:AyLtot_vs_dpdf-}
        }
        \put(-175,70){\rotatebox{90}{$a_{\eta_l}$}}
        \put(-80,-5){$\eta_l^{\textrm{min}}$}
        \put(-80,-15){$$}
      }
      \\
    \end{tabular}
    \caption{Pseudorapidity asymmetry $a_{\eta_l}$ for $pp$ collisions at
      $\sqrt s=14$ TeV evaluated using different dPDFs.  No cuts are
      applied.}
    \label{fig:AyLtot_vs_dpdf}
  \end{center}
\end{figure}

This quantity is plotted as a function of the minimum pseudorapidity cut on the detector hemispheres in
figure \ref{fig:AyLtot_vs_dpdf}. The GS09 values for $a_{\eta_l}$ are larger because the probability of a proton providing 
two large $x$ (valence) partons is reduced under GS09 (due to number effects), so GS09 results in 
fewer events with the two leptons in the same hemisphere.

In addition to a comparison of GS09 and factorised forms, \cite{Gaunt:2010pi} examines sources of
background aside from the canonical SPS WW processes. It is found that there are several processes
that can give rise to the same-sign lepton signal in the detectors:
\begin{align}   
&\text{Heavy flavour:}&gg \to& t\bar{t},    &t \to& \Wp b \rightarrow l^+\nu b,      &\bar{t}\to& \Wm \bar{b} \rightarrow q\bar q'l^+\nu \bar{c} \\
&                     &gg \rightarrow&\bbbar \to B \bar B + ...,     &B \to& l^+\nu X ,  &\bar{B}^0 \to& B^0 \rightarrow l^+\nu \tilde{X} 
\end{align}
\vspace{-15pt}
\begin{align}
&\text{Electroweak gauge boson pair:} & q\bar{q}' \rightarrow& \Wp \Zgam\rightarrow l^+\nu l^+(l^-) &&&&&&&&& \\
&                                     & q\bar{q}^{\phantom{'}}\rightarrow& \Zgam\Zgam \rightarrow l^+(l^-)l^+(l^-) &&&&&&&&&
\end{align}

The leptons in brackets must fall outside the detector acceptance to give rise to the appropriate signal, 
which in our study meant that they had to satisfy $|\eta| > 2.5$. The variety and strength of the 
backgrounds means that a careful choice of cuts is required to enhance the signal/background as much as 
possible -- these are listed in \cite{Gaunt:2010pi}. Even with these cuts, the backgrounds are 
non-negligible and it is unlikely that a conclusive discrimination between GS09 and factorised forms
will be possible using this channel in the near future. On the other hand, cleaner channels with 
regards to DPS may exist -- we are currently examining double Drell-Yan as a possibility.

\bibliography{jrgdisproc}{}
\bibliographystyle{JHEP}

 
\end{document}